\documentclass[prl,twocolumn,showpacs]{revtex4}
\newcommand{\dket}[1]{| \, #1 \rangle\!\rangle}
\newcommand{\dbra}[1]{\langle\!\langle #1 \, |}

\def\kk{\rangle\!\rangle}

\def\>{\rangle}
\def\<{\langle}

\def\Tr{\hbox{Tr}}
\usepackage{graphicx}
\begin{document}

\title{Information-disturbance tradeoff in estimating a maximally
  entangled state}\author{Massimiliano F. Sacchi} 
\affiliation{QUIT Quantum Information Theory Group}
\homepage{http://www.qubit.it}
\affiliation{CNR - Istituto Nazionale per la Fisica della Materia,
  Dipartimento di Fisica ``A. Volta'',  via A. Bassi 6, I-27100 Pavia, Italy} 
\altaffiliation[Also at ]{CNISM - Consorzio Nazionale
  Interuniversitario per le Scienze Fisiche della Materia}

\date{\today}

\begin{abstract}
  We derive the amount of information retrieved by a quantum
  measurement in estimating an unknown maximally entangled state,
  along with the pertaining disturbance on the state itself.  The
  optimal tradeoff between information and disturbance is obtained,
  and a corresponding optimal measurement is provided.
\end{abstract}

%\pacs{03.65.Wj}
\maketitle

The tradeoff between information retrieved from a quantum measurement
and the disturbance caused on the state of a quantum system is a
fundamental concept of quantum mechanics and has received a lot of
attention in the literature
\cite{WH,wod,sc,sten,eng,fuchs96.pra,banaszek01.prl,fuchs01.pra,banaszek01.pra,
  barnum02.xxx,gmd,ozw,mista05.pra,dema,macca,cv}. Such an issue is
studied for both foundations and its enormous relevance in practice,
in the realm of quantum key distribution and quantum cryptography
\cite{key,key2}.

A part from many heuristic statements of the information-disturbance
tradeoff, just a few quantitative derivations have been obtained in
the scenario of quantum state estimation \cite{hol,qse}. The optimal
tradeoff has been derived in the following cases: in estimating a
single copy of an unknown pure state \cite{banaszek01.prl}, many
copies of identically prepared pure qubits \cite{banaszek01.pra}, a
single copy of a pure state generated by independent phase-shifts
\cite{mista05.pra}, and an unknown coherent state \cite{cv}. Recently,
experiment realization of minimal disturbance measurements has been
also reported \cite{dema,cv}.

The problem is typically the following. One performs a measurement on
a quantum state picked (randomly, or according to an assigned a priori
distribution) from a known set, and evaluates the retrieved
information along with the disturbance caused on the state. The
physical transformation will be described by a quantum operation (in
an old-fashioned terminology, a measurement of the first kind, where
it is possible to describe the state {\em after} the measurement). To
quantify the tradeoff between information and disturbance, one can
adopt two mean fidelities \cite{banaszek01.prl}: the estimation
fidelity $G$, which evaluates on average the best guess we can do of
the original state on the basis of the measurement outcome, and the
operation fidelity $F$, which measures the average resemblance of the
state of the system after the measurement to the original one.

In this Letter, we study and provide the optimal tradeoff between
estimation and operation fidelities when the state is a completely
unknown maximally entangled state of finite-dimensional quantum
systems.  We also provide a measurement that achieves such an optimal
tradeoff. 

The interest in maximally entangled states lies in the fact that they
represent a major resource in quantum information technology, e.g. in
quantum teleportation \cite{telep} and quantum cryptography
\cite{key2}. The study of the information-disturbance tradeoff for
maximally entangled states can become of practical relevance for
posing general limits in information eavesdropping and for analyzing
security of quantum cryptographic communications.

Our results will be obtained by exploiting the group symmetry of the
problem, which allows us to restrict our analysis on {\em covariant
  measurement instruments}. In fact, the property of covariance
generally leads to a striking simplification of problems that may look
intractable, and has been thoroughly used in the context of state and
parameter estimation \cite{hol,qse}.

A measurement process on a quantum state $\rho $ with outcomes $\{r
\}$ is described by an {\em instrument} \cite{instr}, namely a set of
trace-decreasing completely positive (CP) maps $\{{\cal E}_r \}$. Each
map can then be written in the Kraus form \cite{kraus}
\begin{eqnarray}
{\cal E}_r( \rho )= \sum _\mu  A_{r \mu} \rho A_{r \mu} ^\dag \;,
\label{uno}
\end{eqnarray}
and provides the state after the measurement 
\begin{eqnarray}
\rho _r =\frac{{\cal E}_r
(\rho )}{\Tr [{\cal E}_r (\rho )]}
\;,
\end{eqnarray}
along with the probability of
outcome 
\begin{eqnarray}
p_r = \Tr [{\cal E}_r (\rho )] = \Tr\left [\sum _\mu A^\dag _{r
  \mu}A_{r\mu } \rho\right ]
\;.
\end{eqnarray}
The set of positive operators $\{ \Pi _r = \sum _\mu
A^\dag _{r \mu}A_{r\mu }\}$ is known as positive operator-valued
measure (POVM), and normalization requires the completeness relation
$\sum _r \Pi _r = I$. This is equivalent to require that the map $\sum
_r {\cal E}_r $ is trace-preserving. 

\par When considering bipartite systems
it is convenient to exploit the natural isomorphism between operators
$A$ on the Hilbert space $\cal H$ and vectors $\dket{A}$ in ${\cal
  H}^{\otimes 2}$, defined through the equation
\begin{equation}
\dket{A}\equiv \sum_{m,n}\<m|A|n\>|m\>|n\>\,.\label{tre}
\end{equation}
We will make repeated use of the following identities \cite{pla}
\begin{eqnarray}
&&A\otimes B\dket{C}=\dket{ACB^\tau}\,,\\
&& \Tr_1[\dket{A}\dbra{B}]=A^\tau B^*\,,\\
&& \Tr_2[\dket{A}\dbra{B}]=AB^\dag\,,\\
&& \dbra{A}B \rangle \!\rangle =\Tr [A^\dag B]\,,
\end{eqnarray}
where $\tau $ and $*$ denote transposition and complex conjugation
with respect to the fixed basis in Eq. (\ref{tre}), and $\Tr _i$
represents the partial trace over the $i$th Hilbert space.  
A maximally entangled state in ${\cal H}\otimes {\cal H}$, with
$\hbox{dim}({\cal H})=d$ will then be written as $\frac {1}{\sqrt
  d} \dket {U_g}$, where $U_g$ is a unitary $d \times d$ matrix, i.e.
$g$ denotes an element of the group $SU(d)$. When performing averages
on group parameters, for convenience we will take the normalized
invariant Haar measure $dg$ over the group, i.e. $\int_{SU(d)} dg=1$,
and we will also omit $SU(d)$ from the symbol of integral. To avoid
confusion when the number of Hilbert spaces proliferates, we will also
used the notation $\dket{A}_{ij}$ when it is necessary to identify the
vector in the Hilbert space ${\cal H}_i \otimes {\cal H}_j$.
Similarly, $A^{(ij)}$ will denote a linear operator acting on ${\cal
  H}_i \otimes {\cal H}_j$.

\par The operation fidelity $F$ evaluates on average how much the
state after the measurement resembles the original one, in terms of
the squared modulus of the scalar product.  Hence, for a measurement
of an unknown maximally entangled state, one has
\begin{eqnarray}
F = \frac {1}{d^2}  \int dg \sum _{r \mu}  
|\dbra {U_g} A_{r \mu } 
\dket{U_g} |^2 \;,
\end{eqnarray}
where $\{A_{r\mu }\}$ are the Kraus operators of the measurement
instrument (\ref{uno}).  For each measurement outcome $r$, one guesses
a maximally entangled state $\frac {1}{\sqrt d} \dket {U _r}$ and the
corresponding average estimation fidelity is given by 
\begin{eqnarray}
G
%&=&
=
\frac {1} {d^3} \int  dg \sum _{r \mu}  \dbra {U_g} A^\dag _{r\mu}
A_{r \mu}\dket{U_g}  
%\nonumber \\& \times & 
\,|\dbra {U_r} U_g \kk |^2 
\;.\label{gm}
\end{eqnarray}
Without loss of generality, we can restrict out attention to {\em
  covariant} instruments, that satisfy 
\begin{eqnarray}
%&&
{\cal E} _h (U_g \otimes I
\,\rho \, U^\dag _g \otimes I
) 
%\nonumber \\& & 
=(U_g \otimes I ){\cal E}_{g^{-1}h}(\rho )(U^\dag _g \otimes I) 
\;.\label{}
\end{eqnarray}
In fact, for any instrument (\ref{uno}) and guess $\frac {1}{\sqrt d}
\dket {U_r}$ in (\ref{gm}), one can easily show that   
the covariant instrument
\begin{eqnarray}
{\cal E}_h (\rho )  &=& 
\sum _{r \mu} 
(U_h U^\dag _r  \otimes I) A_{r \mu} (U_r U_h^\dag \otimes I)
  \,\rho \,
\nonumber \\& \times & 
(U_h U^\dag _r \otimes I )A_{r \mu} ^\dag  (U_r U_h^\dag
  \otimes I)
\;,\label{cove}
\end{eqnarray}
with continuous outcome $h \in SU(d)$, along with the guess $\frac
{1}{\sqrt {d}}\dket {U_h}$, provides the same values of $F$ and $G$ as
the original instrument (\ref{uno}). 

It is useful now to consider the Jamio\l kowski representation \cite{CJ}, 
that gives a one-to-one correspondence between a CP map ${\cal E}$
from ${\cal H}_{in}$ to ${\cal H}_{out}$ and a
positive operator $R$ on ${\cal H}_{out}\otimes {\cal H}_{in}$ 
through the equations 
\begin{eqnarray}
&&{\cal E}(\rho )=\Tr _{in}[(I_{out} \otimes \rho ^\tau ) R ]\;,
\nonumber \\& & 
R=({\cal E }\otimes I_{in}) \dket{I} \dbra {I} \;.\label{jam}
\end{eqnarray}
When ${\cal E}$ is trace preserving, one has also $\Tr
_{out}[R] = I_{in}$. 

For covariant instruments ${\cal E}_g$ acting on ${\cal H}_1\otimes
{\cal H}_2$ as in Eq. (\ref{cove}), the operator $R_g$
acts on ${\cal H}^{\otimes 4}$, and 
has the form 
\begin{eqnarray}
R_g = U_g ^{(1)}\otimes U_g ^{*(3)} R_0 U_g ^{\dag (1)}\otimes U_g
^{\tau  (3)}
\;,\label{rggg}
\end{eqnarray}
with $R_0 \geq 0$, and the trace-preserving condition 
\begin{eqnarray}
\int dg \, \Tr _{34}[R_g] = I^{(12)}
\;.\label{34}
\end{eqnarray}
From Eq. (\ref{rggg}) and the identity (Schur's lemma for irreducible
group representations \cite{zelo})
\begin{eqnarray}
\int dg \,U_g X
U_g^\dag=\frac 1d \Tr[X] I \label{eq:gravtrc}\;,
\end{eqnarray}
it follows that condition (\ref{34}) is equivalent to 
\begin{eqnarray}
\Tr _{1,3,4}[R_0]=d I^{(2)}  \;, \label{tr134}
\end{eqnarray}
which implies  $\Tr [R_0]=d^2$.

By defining the projector on the unnormalized maximally entangled
vector of ${\cal H}_i \otimes {\cal H}_j$ as 
\begin{eqnarray}
{\cal
  I}^{(ij)}=\dket{I}_{ij}{}_{ij}\dbra{I}
\;,
\end{eqnarray}
the fidelities $F$ and $G$ can be written as
$F=\Tr [R_F R_0]$ and $G=\Tr [R_G R_0]$, where $R_F$ and $R_G$ are the
following positive operators
\begin{eqnarray}
R_F &=& \frac{1}{d^2} 
\int dg \,U_g ^{(1)}\otimes U_g ^{*(3)} \,{\cal I}^{(12)} \otimes
{\cal I}^{(34)} \, U_g ^{\dag (1)}\otimes U_g ^{\tau (3)}
\;,\nonumber \\
%\end{eqnarray}
%and 
%\begin{eqnarray}
R_G&=& \frac{1}{d^3}\int dg \, |\dbra{I} U_g \rangle \!\rangle
|^2\,
%\times \nonumber \\& & 
U_g ^{*(3)} (I^{(12)}\otimes {\cal I}^{(34)}) U_g^{\tau (3)}
\nonumber \\& =&
\frac{1}{d} \{ 
I ^{(12)} \otimes \Tr _{12} [{\cal I}^{(12)} \otimes 
I^{(34)} R_F]\}
\;.\nonumber 
\end{eqnarray}
Using the identity 
(Schur's lemma for reducible group representations \cite{zelo})
\begin{eqnarray}
&&
\int  dg
\,U_g \otimes U_g ^* \,Y \, U_g^{\dag }\otimes U_g ^\tau  =
\Tr[Y {\cal I}/d]{\cal I}/d 
\nonumber \\& & 
+\Tr [Y (I -{\cal I}/d )]\frac{I -{\cal I}/d}{d^2 -1}
\label{eq:grav2}\;,
\end{eqnarray}
%Notice  that $\dket{I}_{12} \dket {I}_{34} =\dket{I}_{13,24}$. 
one obtains 
\begin{eqnarray}
R_F 
&= & 
\frac{1}{d^2 (d^2 -1)} \left [ I + {\cal I}^{(13)}\otimes
{\cal I}^{(24)} 
\right. \nonumber \\
& - & \left.
\frac 1d (I^{(13)}\otimes {\cal I}^{(24)}
+{\cal I}^{(13)}\otimes I^{(24)}
)\right ]
\;,
\nonumber \\
R_G &= &  
\frac{1}{d^2 (d^2 -1)} \left [\left (1 -\frac{2}{d^2}\right )I + \frac 
1d I^{(12)}\otimes
{\cal I}^{(34)}\right ] \;.
\nonumber 
\label{rgg}
\end{eqnarray}
The optimal tradeoff between $F$ and $G$ can be found by looking for a
positive operator $R_0$ that satisfies the trace-preserving condition 
(\ref{tr134}) and 
maximizes a convex combination 
\begin{eqnarray}
p G +(1-p) F = 
 \Tr \{[pR_G +(1-p)R_F ]R_0\} \;, 
\end{eqnarray}
where $p \in [0,1]$ controls the tradeoff between the quality of the
state estimation and the quality of the output replica of the
state. Then, $R_0$ will provide a covariant instrument that
achieves the optimal tradeoff.  It turns out that for any $p$ the
eigenvector corresponding to the maximum eigenvalue of $C(p)\equiv p
R_G +(1-p)R_F$ is of
the form \cite{nota2}
\begin{eqnarray}
|\chi \rangle = x \dket {I }_{12}\dket {I}_{34} + y 
\dket {I }_{13}\dket {I}_{24} 
\;,\label{frm}
\end{eqnarray}
with suitable positive $x$ and $y$. Upon taking $R_0 $ proportional to
$|\chi \rangle \langle \chi |$, the covariant instrument will then be
optimal. In fact, condition (\ref{tr134}) can be easily verified, and
the normalization can be derived from the condition $\Tr [R_0]=d^2$.

From Eqs. (\ref{jam}) and (\ref{rggg}), it follows that the optimal
tradeoff can be reached by an instrument with Kraus operators
\begin{eqnarray}
A_g = a \dket{U_g} \dbra{U_g} +b I\;,\label{agopt}
\end{eqnarray}
where $0\leq a \leq 1$, and $b=\frac 1d (\sqrt{d^2 (1-a ^2)+a^2}-a)$.
In fact, condition $\Tr [R_0]=d^2$ is equivalent to
$(a^2+b^2)d^2+2abd=d^2$.  The corresponding fidelities are given by
\begin{eqnarray}
F&=&\frac{1}{d^2(d^2 -1)}[d^2 +(d^2 -2)(a+b d)^2] 
%\nonumber \\&= & 
= 1
- \frac{d^2 -2}{d^2}a^2\;, \nonumber \\
 G &=& \frac{1}{d^2(d^2 -1)}[d^2 -2 + (ad+b)^2]
=\frac{2- b^2}{d^2}\;.\nonumber 
\end{eqnarray}
Notice that the instrument given by operators (\ref{agopt}) is {\em
  pure}, in the sense that it leaves pure states as pure.  When no
measurement is performed ($a=0$), one has $F=1$ and $G=\frac
{1}{d^2}$, which is equivalent to randomly guessing the unknown state.
The optimal estimation can be obtained by a Bell measurement ($b=0$),
namely by projectors on maximally entangled states, and gives
$F=G=\frac{2}{d^2}$.

Upon eliminating $a$ and $b$, we obtain the optimal tradeoff between
$F$ and $G$ 
\begin{eqnarray}
\sqrt{(d^2-2)(2- d^2 G)} = \sqrt{(d^2 -1)F -1} -\sqrt{1-F}\;,\label{gf}
\end{eqnarray}
or, equivalently,
\begin{eqnarray}
&&
\label{fg}
\sqrt{\frac{d^2}{d^2-2}\left(F -\frac{1}{d^2 -1}\right )}
\\& & 
= \sqrt{G-\frac{d^2 -
2}{d^2(d^2 -1)}} +\sqrt{(d^2-1)\left (\frac {2}{d^2}-G \right
)}\;.
\nonumber 
\end{eqnarray}
%Equations (\ref{gf}) and (\ref{fg}) with equality gives the optimal
%tradeoff, and for any suboptimal measurement strict inequalities hold.
Such an optimal tradeoff overcomes the corresponding one for a
completely unknown state \cite{banaszek01.prl} in a Hilbert space with
dimension $d^2$, i.e.  for a fixed value of the estimation fidelity
$G$ one can achieve here a better value of the operation fidelity
$F$. In other words, when a partial knowledge of the set of states is
available (here, the fact that the states are maximally entangled), one
can obtain the same estimation fidelity with a smaller disturbance of
the state. 

We can introduce two normalized quantities---a sort of
visibilities---that can be interpreted as the average information $I$
retrieved from the quantum measurement and the average disturbance $D$
affecting the original quantum state as follows:
\begin{eqnarray}
I=\frac{G - G_0}{G_{max}-G_0}=d^2G -1=1-b^2
\;,\label{igg}
\end{eqnarray}
where $G_0=\frac {1}{d^2}$ is the value of $G$ for random guess and
$G_{max}=\frac{2}{d^2}$ is the maximum value attainable by $G$; 
\begin{eqnarray}
D= \frac{1-F}{1-F_{min}}=\frac{d^2(1-F)}{d^2 -2}=a^2
\;,\label{dgg}
\end{eqnarray}
where $F_{min}=\frac {2}{d^2}$ represents the average fidelity with
the maximally chaotic state $\frac {I}{d^2}$.  Clearly, one has $0\leq
I \leq 1$, and $0\leq D \leq 1$.  In this way, after some algebra one
obtains the quadratic expression
\begin{eqnarray}
d^2(D-I)^2 -4D (1-I)=0\;
\end{eqnarray}
that gives the optimal information-disturbance tradeoff. We plot in
Fig. 1 the behavior of the tradeoff for dimension $d=2,4$, and $8$.
For a given value of the information $I$, the curves $D(I)$ represent
a lower bound for the disturbance of any measurement instrument.

\begin{figure}[htb]
\begin{center}
\includegraphics[scale=.8]{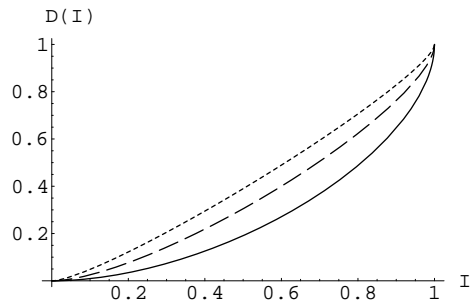}
\caption{Optimal information-disturbance tradeoff in estimating an
  unknown maximally entangled state for dimension $d=2$ (solid line),
  $d=4$ (dashed line), and $d=8$ (dotted), where $I$ and $D$ are
  defined through Eqs. (\ref{igg}) and (\ref{dgg}) in terms of the
  estimation and operation fidelities $G$ and $F$, respectively. For
  given value of the retrieved information $I$, the curves $D(I)$ are
  a lower bound for the disturbance of any measurement instrument.}
\label{f:fig1}
\end{center}
\end{figure}

The optimal tradeoff is reached by a measuring instrument
(\ref{agopt}) whose Kraus operator are a coherent superpositions of two
extreme measurements: the identity map and the optimal map for
estimating an unknown maximally entangled state.  It can be easily
shown that the discrete version $\{{\cal E}_r \}$ of such an
instrument with Kraus operators
\begin{eqnarray}
A_r = \frac{1}{d}(a \dket {U_r}\dbra{U_r} +b I )\;, \qquad r=1,2,...,d^2 
\;,
\end{eqnarray}
and orthogonal $\{ \dket{U_r} \}$, namely $\dbra{U_r} U_s \rangle
\!\rangle =d \delta _{rs}$, achieves the same values of $F$ and $G$,
and hence the optimal tradeoff as well. Notice that the POVM $A^\dag
_r A_r$ corresponding to this instrument is made of projectors on
so-called Werner states \cite{ws}, i.e. convex mixtures of maximally
entangled and maximally chaotic states. The experimental realization
of such a kind of measurement could be investigated for hyperentangled
two-photon states, for which Bell measurements have been already
demonstrated \cite{padua}. 

\par In conclusion, a tight bound between the quality of estimation of
an unknown maximally entangled state and the degree the initial state
has to be changed by this operation has been derived. Such a bound can
be achieved by noisy Bell measurements, where the noise continuously 
controls the tradeoff between the information retrieved by the measurement
and the disturbance on the original state. 

{\em Acknowledgments.} This work has been sponsored by Ministero
Italiano dell'Universit\`a e della Ricerca (MIUR) through FIRB (2001)
and PRIN 2005.


\begin{thebibliography}{99}
\bibitem{nota2}It can be shown analytically that the operator norm
 of $C(p)$ (i.e. its maximum eigenvalue) satisfies 
\begin{eqnarray}
&&
\| C(p) \|=
  \frac{1}{2d^4(d^2-1)}\{d^4-4p+3d^2 p -d^4 p 
\nonumber \\& & 
+d \sqrt{d^2[2-d^2(1-p)-p]^2 +4(d^2-2)p(1-p)}\}
\nonumber 
\end{eqnarray}
for $d=2,3$, and the
  corresponding eigenvector is of the form as in Eq. (\ref{frm}), with
  $x=1$ and 
\begin{eqnarray}
&&
y=\frac {1}{2p}
\{d^3(1-p)-d(2-p)+ 
\nonumber \\& & 
\sqrt{d^2[2-d^2(1-p)-p]^2
    +4(d^2-2)p(1-p)}\} \nonumber 
\end{eqnarray}
For higher values of $d$, such a result can be checked, for example,
by means of the power method \cite{pm}.
\end{thebibliography}
\end{document}